\documentclass[12pt]{iopart}

\usepackage{iopams}  
\usepackage{bm}
\usepackage{graphicx}
\usepackage{cite}
\begin{document}

\title{Fracturing of topological Maxwell lattices
}

\author{Leyou Zhang \& Xiaoming Mao}

\address{Department of Physics, University of Michigan, 450 Church Street, Ann Arbor, Michigan, MI, 48109}
%\ead{submissions@iop.org}
\vspace{10pt}
\begin{indented}
\item[]January 2018
\end{indented}

\begin{abstract}
We present fracturing analysis of topological Maxwell lattices when they are stretched by applied stress.  Maxwell lattices are mechanical structures containing equal numbers of degrees of freedom and constraints in the bulk and are thus on the verge of mechanical instability.  Recent progress in topological mechanics led to the discovery of topologically protected floppy modes and states of self stress at edges and domain walls of Maxwell lattices.  When normal brittle materials are being stretched, stress focuses on crack tips, leading to catastrophic failure.  In contrast, we find that when topological Maxwell lattices are being stretched, stress focuses on states of self stress domain walls instead, and bond-breaking events start at these domain walls, even in presence of cracks.  Remarkably, we find that the stress-focusing feature of the self-stress domain walls persists deep into the the failure process, when a lot of damages already occurred at these domain walls.  We explain the results using topological mechanics theory and discuss the potential use of these topological Maxwell lattice structures as mechanical metamaterials that exhibit high strength against fracturing and well controlled fracturing process.

\end{abstract}

% Uncomment for PACS numbers
%\pacs{00.00, 20.00, 42.10}
%
% Uncomment for keywords
%\vspace{2pc}
%\noindent{\it Keywords}: XXXXXX, YYYYYYYY, ZZZZZZZZZ
%
% Uncomment for Submitted to journal title message
%\submitto{\JPA}
%
% Uncomment if a separate title page is required
%\maketitle
% 
% For two-column output uncomment the next line and choose [10pt] rather than [12pt] in the \documentclass declaration
%\ioptwocol
%

\section{Introduction: topological mechanics and Maxwell lattices}

In recent years, there has been substantial advances in applying the conceptual framework of topological states of matter to classical mechanical systems which are governed by Newton's laws \cite{kane2014topological,Prodan2009,sun2012surface,Chen2014,Wang2015,nash2015topological,chen2016topological,zhou2017topological,
paulose2015topological,susstrunk2015observation,mousavi2015topologically,Yang2015,
Peano2015,Strohm2005,Sheng2006,Pal2016,He2016,Suesstrunk2016,
Xiao2015a,rocklin2016mechanical,rocklin2017transformable,paulose2015selective}. These mechanical systems can acquire exotic mechanical behaviors, such as one-way wave transport  \cite{nash2015topological,Wang2015,susstrunk2015observation,mousavi2015topologically,Yang2015,
Peano2015,Strohm2005,Sheng2006,Pal2016,He2016}, nonlinear soliton~\cite{Chen2014}, switchable stiffness~\cite{rocklin2017transformable},  and selective buckling \cite{paulose2015selective}, that originate in the topological states of their phonon band structures.
 Many of these systems belong to the class of Maxwell lattices, lattices that contain equal number of degrees of freedom and constraints in the bulk and hence are on the verge of mechanical instability. For any ideal mechanical system that consists mass points connected by harmonic springs on a periodic lattice in $d$-dimensional space, the condition for a Maxwell lattice is $\langle z \rangle=2d$, where $\langle z \rangle$ is the mean coordination number~\cite{lubensky2015phonons,Souslov2009,Mao2010,Ellenbroek2011,Mao2011a,sun2012surface,Mao2013b,
Mao2013c,Zhang2015a,Mao2015}.  This condition comes from balancing the degrees of freedom per site, $d$, with the average number of constraints per site, $\langle z \rangle/2$.  

For general mechanical structures with point-like particles and central-force bonds one can apply the Maxwell-Calladine counting rule~\cite{Calladine1978,sun2012surface,kane2014topological,lubensky2015phonons}, 
\begin{equation}
N_0 - N_{SS} = N_s d - N_b,
\end{equation}
where $N_s$ is the number of sites in the lattice, $N_b$ is the number of springs, $N_0$ is the counting of floppy modes (modes of zero elastic energy) and $N_{SS}$ is the counting of states of self stress (eigenstates of tension and compression on bonds with no net force on any site).  
For a periodic Maxwell lattice which has no boundaries, we always have $N_s d - N_b = 0$ and therefore $N_0=N_{SS}$. 
A finite Maxwell lattice has fewer springs on the lattice boundaries and therefore $N_s d - N_b > 0$ and $N_0>N_{SS}$ so there must be floppy modes. While the Maxwell-Calladine counting rule give us the relation between $N_0$ and $N_{SS}$, it doesn't tell us where the floppy modes and states of self stress are in space, and whether they are localized or extended in the Maxwell lattice, which depend on the actual lattice architecture.  A topological invariant, $\bm R_{\rm T}$, called ``topological polarization'', was introduced by Kane and Lubensky in Ref.~\cite{kane2014topological}, to characterize localization of floppy modes and states of self stress.  We will discuss more details of the topological polarization in the Theory section below.  
Similar to the topological states of matter in other systems, mechanical behaviors that are rooted in this topological invariant remain robust even when the system is locally perturbed. This makes topological mechanics a promising approach for designing mechanical  metamaterials that are insensitive to impurities and defects.

In this paper, we study how Maxwell lattices fracture when they are macroscopically deformed by a uniaxial strain applied to the boundaries. We are especially interested in the fracturing process of lattices that are topologically polarized, i.e. $\bm R_{\rm T}\neq 0$, and contain topologically protected localized states of self stress. 
It was shown in Ref.~\cite{paulose2015selective} that when Maxwell lattices with self-stress domain walls buckle under pressure, the buckling regions localize to the self-stress domain walls because stress focuses on these domain walls.  Similarly, we find that  fracturing of these Maxwell lattices under external tension starts at these self-stress domain walls due to this stress-focusing effect.  More interestingly, even after a significant part of the self-stress domain walls is damaged during the fracturing process, stress still robustly focuses to these domain walls.  After the first bond-breaking event, the lattice mean coordination number decreases below the Maxwell point, $\left<z\right><2d$, and thus simple topological mechanics theory that predicts localized states of self stress no longer directly apply.  The fact that stress still focuses on the self-stress domain walls when the lattice is damaged originates from the robustness of these topologically protected edge states of these Maxwell lattices.

The stress and damage focusing effect of self-stress domain walls in Maxwell lattices provides remarkable protection on the rest of the structure.  As we show below, these self-stress domain walls guide stress away from preexisting cracks in the structure.  Thus, unlike normal materials where stress focuses on crack tips leading to catastrophic failure~\cite{Griffith1921}, Maxwell lattices with self-stress domain walls exhibit a much more gradual fracturing process where only a small number of bonds break at each step of strain increase, without any large avalanches.  These observations reveal the great potential of using Maxwell lattices to design high-strength metamaterials in which failure occurs gradually at predicted locations.

\begin{figure}[htbp]
\begin{center}
\includegraphics{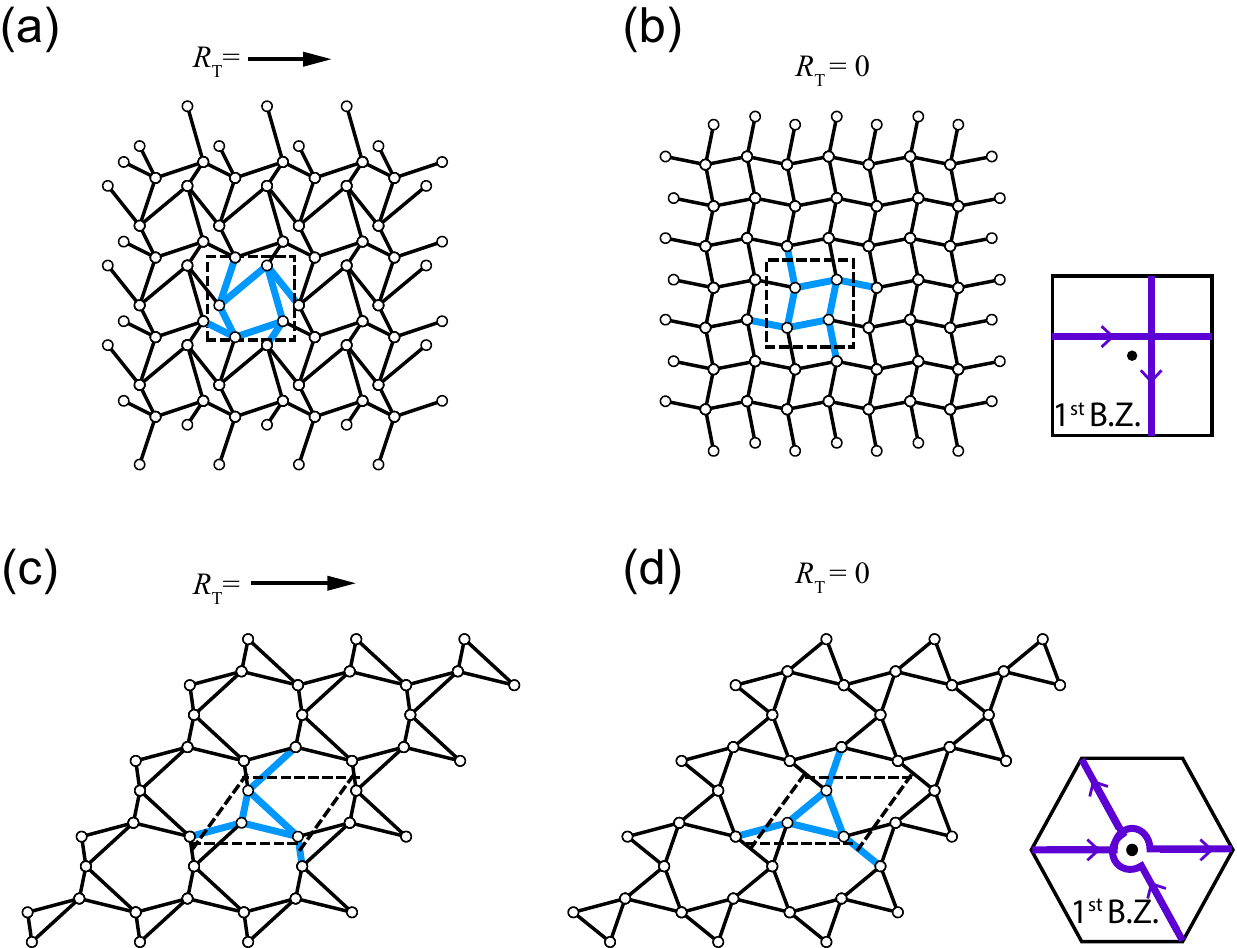}
\caption{Deformed square (a,b) and deformed kagome (c,d) lattices that are used in the simulation of lattice fracturing. For both types of lattice, we choose one lattice that is topologically polarized (a,c) and one lattice that is not (b,d). The positions of the lattice sites inside one unit cell of the lattices are (a) $\bm{r}_1=(0,0),\bm{r}_2=(0.6,-0.2),\bm{r}_3=(0.4,-0.9),\bm{r}_4=(-0.2,-0.4)$, (b) $\bm{r}_1=(0,0),\bm{r}_2=(0.5,-0.1),\bm{r}_3=(0.6,-0.6),\bm{r}_4=(0.1,-0.5)$, (c) $\bm{r}_1=(0,0),\bm{r}_2=(0.52,0.13),\bm{r}_3=(0.06,-0.3)$ and (d) $\bm{r}_1=(0,0),\bm{r}_2=(0.52,0.13),\bm{r}_3=(0.36,-0.3)$, and the primitive vectors of the lattices are (a), (b) $\bm{e}_1=(1,0), \bm{e}_2=(0,1)$ and (c), (d) $\bm{e}_1=(1,0), \bm{e}_2=(1/2,\sqrt3/2)$. In each of the four lattices, unit cells are marked by the black dotted boundary, and bonds that belong to the marked unit cell are marked as thick blue solid lines (other bonds in the lattice are thin black solid lines).  Each unit cell in the deformed square lattice contains four sites, and each unit cell in the deformed kagome lattice contains three sites.
The integral paths for determining $\bm R_{\rm T}$ are also marked in the first Brillouin zone (insets on the right) of the lattices.  }
\label{fig1}
\end{center}
\end{figure}

\begin{figure}[htbp]
\begin{center}
\includegraphics[width=0.8\textwidth]{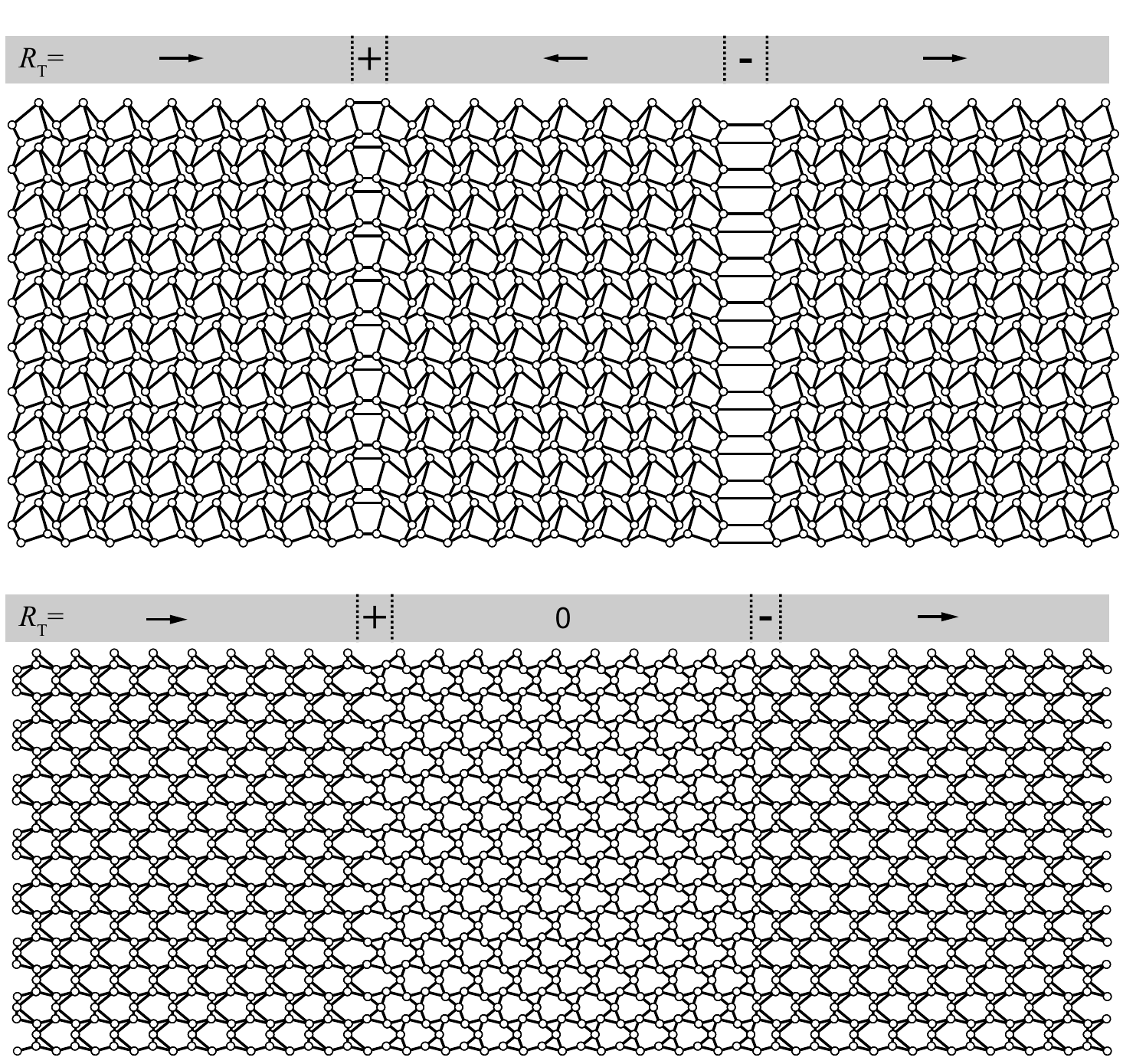}
\caption{The deformed square (top) and deformed kagome (bottom) lattices with domain walls that have $\nu^{\mathcal{S}}_{\rm T}\neq 0$. Lattice-domain topological polarizations $\bm R_{\rm T}$ are marked in the bar on top of each lattice. 
There are floppy modes that are localized on the domain wall with $\nu^{\mathcal{S}}_{\rm T}> 0$ (marked ``$+$'' in the bar),  and  self stress on the domain wall with $\nu^{\mathcal{S}}_{\rm T} < 0$ (marked ``$-$'' in the bar). 
}
\label{fig2}
\end{center}
\end{figure}

To study the fracturing of Maxwell lattices, we choose two types of two-dimensional (2D) Maxwell lattices, the deformed square lattice and deformed kagome lattice, which are commonly used in the research of topological mechanics \cite{kane2014topological,lubensky2015phonons,rocklin2016mechanical}. Both types of these Maxwell lattices can acquire nonzero $\bm R_{\rm T}$ by varying site positions in the unit cell. We  first generate ``phase diagrams'' of $\bm R_{\rm T}$ in the parameter space of site positions and then determine the site positions that we need to polarize the lattice with desired $\bm R_{\rm T}$. In fig.~\ref{fig1}, we show these lattices with their topological polarization $\bm R_{\rm T}$. The configurations of their unit cells will be used throughout this paper. To acquire states of self stress that are protected by topology, we introduce multiple domains of these lattices with different $\bm R_{\rm T}$, which are separated by domain walls of floppy modes and states of self stress. Such structures  are shown in fig.~\ref{fig2} for both the deformed square and the deformed kagome lattices.  The connection between $\bm R_{\rm T}$, domain walls and protected states of self stress will be explained in Sec.~\ref{SEC:Theory}.

\section{Theory: topological polarization and exponential decay of  topological surface modes}\label{SEC:Theory}

In Maxwell lattices, the coordination number $\langle z \rangle=2d$ ensures that the counting of constraints and degrees of freedom always equal locally, except near open boundaries of the lattices, where there are fewer constraints than degrees of freedom, giving rise to floppy modes.  Whether an edge of a Maxwell lattice hosts exponentially localized floppy modes is characterized by the topological polarization $\bm{R}_{\rm T}$, which is defined through calculating  winding numbers of the determinate of the compatibility matrix $C(q)$ over closed paths $\mathcal{C}$ (fig.~\ref{fig1}) across the first Brillouin zone, i.e.~\cite{kane2014topological},
\begin{equation}
n = \frac{1}{2\pi i}\oint_\mathcal{C}dq \frac{d}{dq}\ln \det C(q).
\end{equation}
The compatibility matrix $C(q)$, as we discuss below in this Section, is a linear operator that maps lattice site displacements to bond extensions.  
We follow two paths along the reciprocal lattice vectors $\bm{q}_1$ and $\bm{q}_2$ and calculate the topological winding numbers $n_1$ and $n_2$ for any 2D periodic lattice. The topological polarization for a periodic Maxwell lattice is then defined as $\bm{R}_{\rm T} = n_1\bm{r}_1+n_2\bm{r}_2$, where $\{\bm{r}_i\}$ are the real space primitive vectors that correspond to the reciprocal vectors $\{\bm{q}_i\}$ in momentum space.

To determine which boundaries have exponentially localized floppy modes or states of self stress, we need to calculate the surface integral of $\bm{R}_{\rm T}$ over a surface (contour for 2D lattices) $\mathcal{S}$ that encloses the lattice boundary of interest,.i.e.~\cite{kane2014topological}, 
\begin{equation}
\nu^{\mathcal{S}} _{\rm T} = \frac{1}{V_0}\oint_\mathcal{S} \bm{R}_{\rm T}\cdot \bm{G} ds,
\end{equation}
where $V_0$ is the volume of the unit cell, $\bm{G}$ is the reciprocal vector orthogonal to the surface $\mathcal{S}$.
With symmetric choice of unit cells where the number of cut bonds are the same on opposite edges (this is characterized by the local count $\nu^{S}_{\rm L}$ as defined in Ref.~\cite{kane2014topological} and symmetric unit cell means $\nu^{S}_{\rm L}=0$), $\nu^{\mathcal{S}}_{\rm T}$ directly gives the difference between the numbers of floppy modes and states of self stress inside the surface $\mathcal{S}$.    
In this paper we construct the deformed square and deformed kagome lattices with such symmetric unit cells.  As a result, 
  $\nu^{\mathcal{S}} \equiv N_0 - N_{SS} = \nu^{\mathcal{S}}_{\rm T}+\nu^{\mathcal{S}}_{\rm L}=\nu^{\mathcal{S}}_{\rm T}$ is the counting of how many more localized floppy modes are inside $\mathcal{S}$ than localized states of self stress. Since both $N_0$ and $N_{SS}$ are non-negative, when $\nu^{\mathcal{S}}>0$, there are guaranteed floppy modes inside $\mathcal{S}$ that are protected by the topology of the bulk phonon structure of the lattice; When $\nu^{\mathcal{S}}<0$, there are guaranteed states of self stress inside $\mathcal{S}$.   

The formulation described above applies equally to lattice surfaces (outer boundaries, with vacuum considered $\bm{R}_{\rm T}=0$) and domain walls (inner boundaries between lattice domains with different $\bm{R}_{\rm T}$).  
Examples of lattices with domain walls are shown in fig.~\ref{fig2}. When $\nu^{\mathcal{S}}>0 (<0)$ for a surface that encloses a domain wall, there will be topologically protected floppy modes (states of self stress) localized on the domain wall.  In these lattices, because the two domains separated by the domain wall are of homogeneous (but different) $\bm{R}_{\rm T}$'s, the value of $\nu^{\mathcal{S}}$ is proportional to the height of the domain wall.

Floppy modes and states of self stress that are localized on boundaries exponentially decay into the lattice bulk because their wave vectors $\bm{q}$ have nonzero imaginary parts, i.e. $\bm{q} = \bm{k} +i\bm{\kappa}$. For example, a floppy modes on a lattice boundary $\bm{u} = \bm{A}\exp{(i\bm{q}\cdot\bm{r})} = \bm{A}\exp{(i\bm{k}\cdot\bm{r}-\bm{\kappa}\cdot\bm{r})}$ decays into the lattice bulk with a decay rate $|\bm{\kappa}|$. The sign of $\kappa$ is dictated by the topological polarization $\bm{R}_{\rm T}$ (which is why $\bm{R}_{\rm T}$ determines which edge or domain wall floppy modes and states of self stress localizes to), whereas the magnitude of $\kappa$ depends on the detailed geometry of the unit cell.  To ensure localization of floppy modes or states of self stress on edges or domain walls, the lattice domain depth has to be greater than $1/|\bm{\kappa}|$.  
To determine the value of $\bm\kappa$, we need to solve $\bm{q}$ from the condition for all floppy modes using the compatibility matrix, 
\begin{equation}
\det C(\bm{q}) = 0.
\end{equation}
Taking the component of $\bm{q}$ along the boundary to be any real wave number, we can solve for $\bm\kappa$ that controls the decay rate perpendicular to the boundary.  
The decay rate of states of self stress can be similarly solved by requiring 
\begin{equation}
\det Q(\bm{q}) = 0,
\end{equation}
where $Q = C^T$ is the equilibrium matrix of the lattice. Since $\det Q = \det C$, the decay rate of states of self stress is the same as the decay rate of the floppy modes in a given domain.

Finally, we need to connect the states of self stress to the linear response of the lattice to macroscopic deformations. We start with the equilibrium matrix $Q$ and compatibility matrix $C$ of a lattice~\cite{PellegrinoCal1986,Calladine1978} 
\begin{eqnarray}
\bm f &= Q\bm t ,\\
\bm e &= C\bm u,
\end{eqnarray}
where $\bm t$ is the force on each springs, $\bm u$ is the displacement of each lattices sites, $\bm f$ is the net force on each lattice sites and $\bm e$ is the extension on each springs. It is straightforward to show that $Q=C^{\rm T}$.  In addition, $\bm t$ is related to $\bm e$ as $\bm t = K \bm e$ where $K$ is the diagonal matrix of the spring stiffness. 

For the study of lattice fracturing under external stress, we  make explicit distinction between lattice boundaries with controlled displacement (denoted as $\partial V$ in subscript) and the lattice bulk ($V$) so that we can impose deformations on the boundaries and equilibrate the lattice in the bulk. The linear equations then become
\begin{eqnarray}
\bm f_{V} &= Q_{V}\bm t ,\\
\bm e &= C_{V}\bm u_{V} + C_{\partial V}\bm u_{\partial V},\label{extension}
\end{eqnarray}
where boundary displacements $u_{\partial V}$ are given.  
We can solve $\bm t$ by setting $\bm f_V=0$ (force balance on all internal sites) and simplify the above equations as
\begin{equation}
Q_{V}\bm t = Q_{V}K(C_{V}\bm u_{V} + C_{\partial V}\bm u_{\partial V}) = 0.
\end{equation}
Therefore $\bm t$ is a null vector of $Q_{V}$ and thus must be a superposition of states of self stress in the lattice bulk, i.e. $\bm t = \sum_i a_i \bm s_i$, where $\{\bm s_i\}$ is the complete orthonormal basis for the null space of $Q_V$. We can further express $\bm t$ as 
\begin{equation}
\bm t = \sum_{i,j} (\bm s_i\cdot K^{-1}\bm s_j)(\bm s_j\cdot C_{\partial V}\bm u_{\partial V})\bm s_i,
\label{tension}
\end{equation}
which shows that when bond extension caused by the boundary deformation $\bm u_{\partial V}$ has nonzero overlap with states of self stress in the bulk, bonds in the lattice will have nonzero tension originated from these bulk states of self stress. We show the derivation of eq.~(\ref{tension}) in detail in the Appendix.

Up to now we haven't consider the possibility that the lattices may have topologically protected Weyl points that can complicate our definition for $\bm R_{\rm T}$. It has been shown that a deformed square lattice with four lattice sites in a unit cell can have zero or two Weyl points in the first Brillouin zone, depending on geometry of the lattice \cite{rocklin2016mechanical}. For a kagome lattice with three sites in the unit cell, there are no Weyl points.  This paper is only concerned with fracturing of lattices with no Weyl points.

\section{Simulation: lattice fracturing due to of external load}\label{SEC:Simu}

In simulation, we build up our system by connecting nearest neighbor (NN) point masses with harmonic springs that are free to rotate around both of their joints. The hamiltonian of our system is therefore 
\begin{equation}
H = \sum_{i,j\in NN} \frac{1}{2}k(|\bm{r}_i-\bm{r}_j|-l_{ij})^2, 
\end{equation}
where $k$ is the spring stiffness which is set to one in the simulation, $\bm{r}_i$ is the position of site $i$ and $l_{i,j}$ is the rest length of the spring between site $i$ and site $j$. The geometry of sites and springs are shown in fig.\ref{fig1} and fig.\ref{fig2} for both the deformed square and the deformed kagome lattices. 

To simulate the fracturing process of the lattice, we impose uniaxial strain in the vertical direction to the lattice boundary by holding the lattice sites on the bottom boundary  stationary vertically while displacing the lattices sites on the top boundary with a distance $\delta h = h \gamma$ up, while $h$ is the height of the lattice and $\gamma$ is the strain that we impose. The lattice sites on top and bottom boundaries are allowed to slide along the horizontal directions of the boundaries. The lattices have a width $w$ and open boundary conditions on the left and the right boundaries. With the boundary condition defined above, we relax the lattice to an energy minimum such that the force of springs are balanced on all internal lattice sites~\cite{bitzek2006structural}.  Thus, the vertical degrees of freedom of the top- and bottom- boundary sites correspond to $\partial V$, and their horizontal degrees of freedom, along with all internal sites correspond to $V$ in our discussion of the theory.  

At every strain step (given $\gamma$), we examine force on springs and compare the amplitude of the force to a compressive strength $f_c$ and a tensile strength $f_t$. Both $f_t$ and $f_c$ define the limit force that a spring can bear, beyond which the spring breaks and is removed from the lattice. When we need to break and remove springs after relaxation, we do so and retake the relaxation method for the new lattice and further remove springs that are beyond compressive/tensile limits.  We repeat this process until force balance is reached with all springs within limits.  We then proceed with the next strain step.  
We set $f_t = f_c\equiv f_0=10^{-2}\ll1$ in our simulation for all springs in the lattices.

\section{Results}

\begin{figure}[htbp]
\begin{center}
\includegraphics[width=1.0\textwidth]{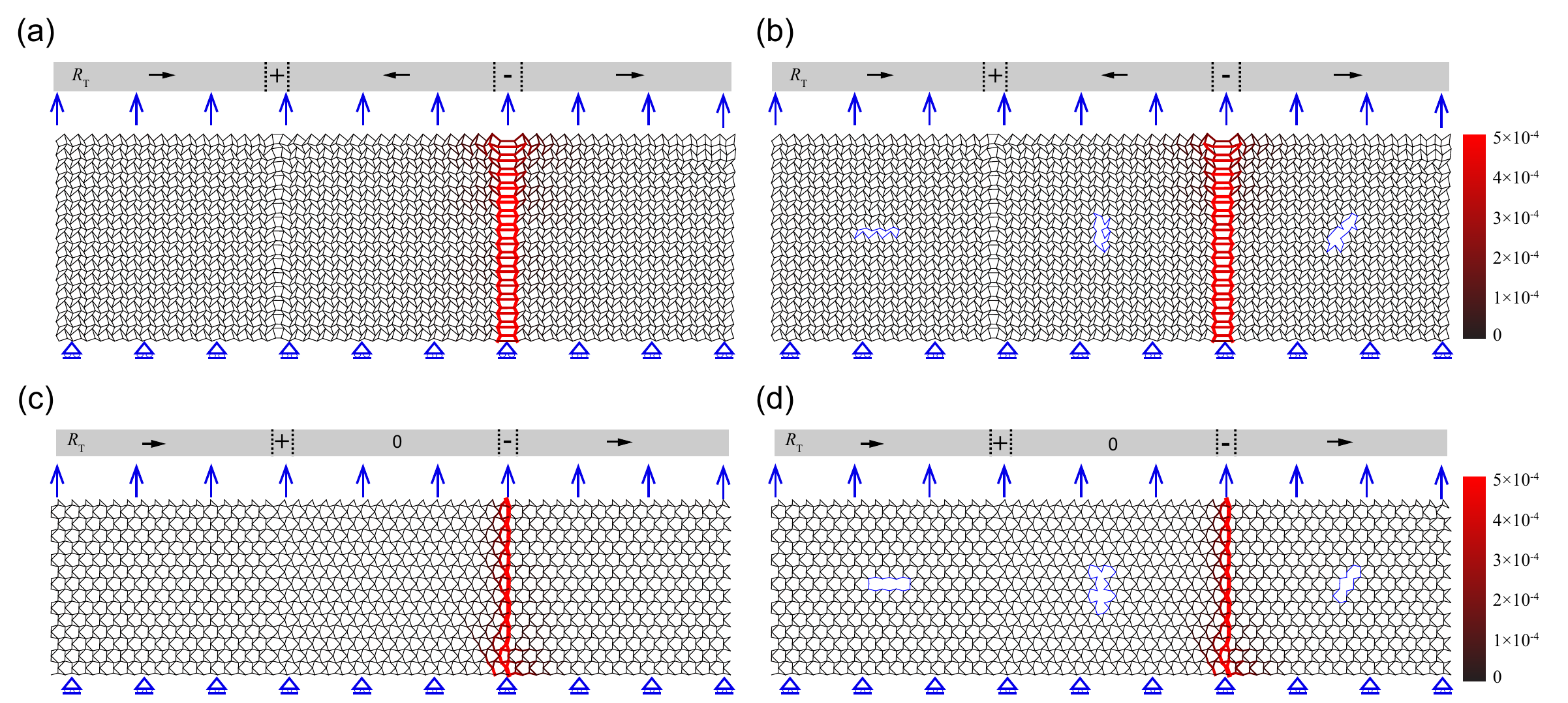}
\caption{The linear response of lattices when a small strain $\gamma=10^{-3}\ll 1$ is imposed in the vertical direction to the deformed square (a,b) and deformed kagome (c,d) lattices. Springs are colored according to the values of their tension (legend on the right). (a) and (c) show lattices with intact bulk domains.  (c) and (d) show lattices with small cracks of pre-removed bonds (highlighted blue on their boundaries) in bulk domains.  In all cases, stress focuses on self-stress domain walls. We mark $\bm R_{\rm T}$ for the bulk of the lattice domains and the sign of $\nu^{\mathcal{S}}$ for domain walls above each lattice.
}
\label{fig3}
\end{center}
\end{figure}

We begin by testing the lattices' linear response to uniaxial load, as shown in fig.~\ref{fig3}.  The simulation protocol is discussed in Sec.~\ref{SEC:Simu}, and to obtain linear response, we use very small strain ($\gamma=10^{-3}\ll 1$) so no bonds are beyond compression/tensile force limit.
We not only include perfect lattices as shown in fig.\ref{fig2} but also include lattices that have cracks in the bulk. 

In fig.~\ref{fig3}, we show the stress distribution of linear response for the lattices. Consistent with theory, high stress appears at the domain walls on the right of the lattice, where the topological polarization ensures a topological charge of $\nu^{\mathcal{S}}\propto -h<0$, an indication that the domain wall acquires localized states of self stress that are protected by the lattice topology. There is no stress localization on the left domain wall, where the topological polarization gives a topological charge of $\nu^{\mathcal{S}}\propto h>0$, indicating protected floppy modes instead of localized states of self stress. 
We also find that unlike normal materials, in which stress is amplified at crack tips~\cite{Griffith1921,dowling2012mechanical}, in these topological Maxwell lattices there is no obvious stress at the tips of the cracks.  Thus, these cracks are protected by the self-stress domain walls. 
In addition, we observe no significant difference of stress distribution between the deformed square and deformed kagome lattices, and between lattices with and without cracks. The overall stress distribution in the linear regime is mainly determined by the topology of the lattices bulk, regardless of the microscopic details and small defects that may exist in the lattices.

We then test the entire process of fracturing of the lattices in the quasi-static limit, where kinetic energy of the lattices is quickly dissipated, much faster than the relaxation process of the lattices.  We increase the strain by small steps such that at each new strain at most one bond breaks initially (there may be subsequent avalanches when we recalculate the stress distribution after the initial bond breaking but in our observation all avalanches are small events with number of breaking bonds of $\mathcal{O}(1)$).  For all lattices shown in fig.~\ref{fig3}, we find that the fracture begins in one of the bonds at the self-stress domain wall where stress localizes, as one expect from the linear theory.  

\begin{figure}[htbp]
\begin{center}
\includegraphics[width=\textwidth]{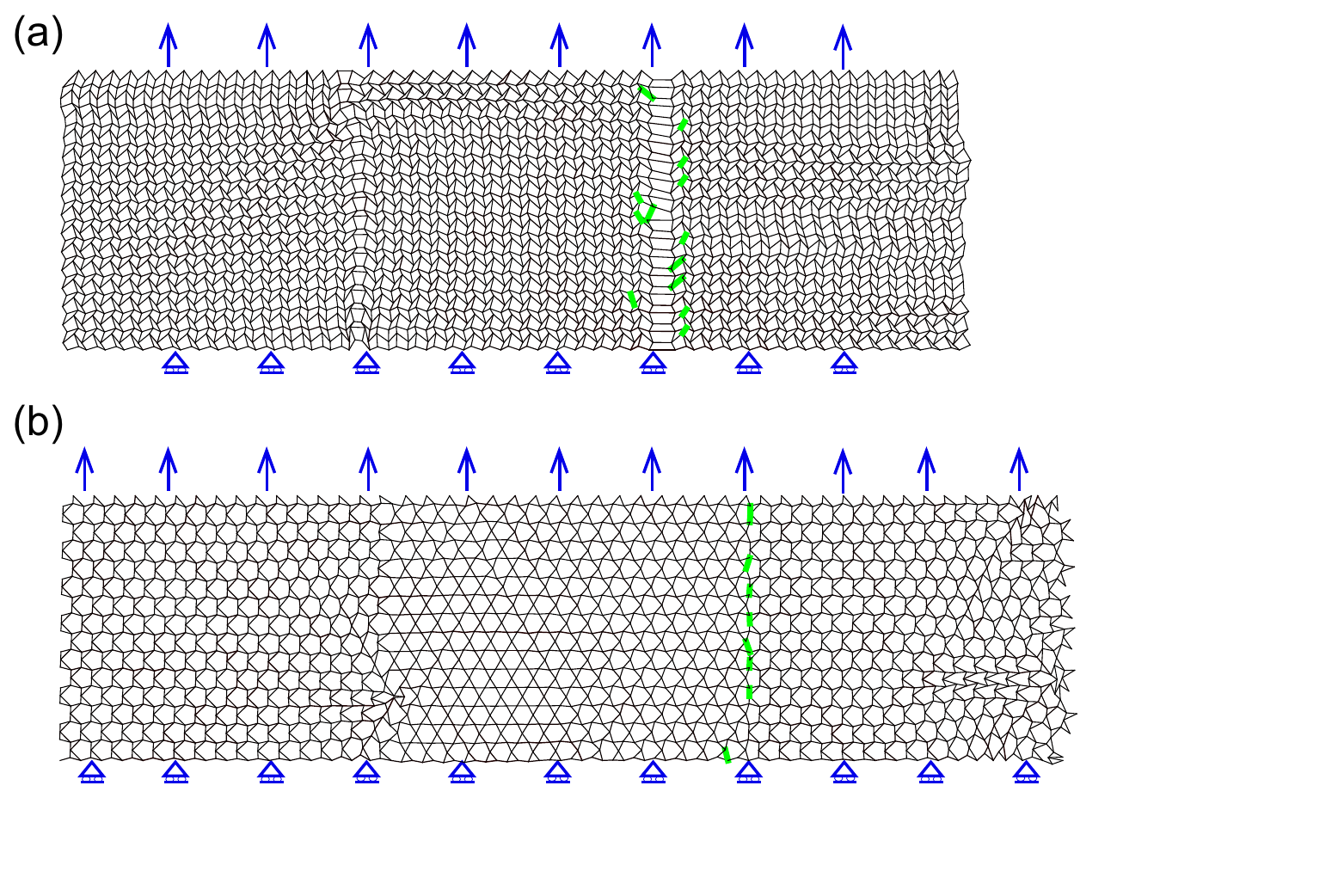}
\caption{
Snapshots of deformed square (a) and deformed kagome (b) lattices that are in the process of fracturing. Broken springs that have been removed from the lattices are marked in green. The broken springs are  localized near the self-stress domain walls with $\nu^{\mathcal{S}}<0$.
}
\label{brokenbonds}
\end{center}
\end{figure}

Interestingly, the next bond breaking events continue to concentrate near the self-stress domain walls (fig.~\ref{brokenbonds}), even when the lattice is rather deep into the failure process and the coordination number becomes $\langle z \rangle <2d$.  This is a manifestation of the robustness of topological protection -- with small damages at the domain wall, the phonon band topology of the bulk of the lattice is unchanged and still dictates the exponential localization of states of self stress and floppy modes.

\begin{figure}[htbp]
\begin{center}
\includegraphics{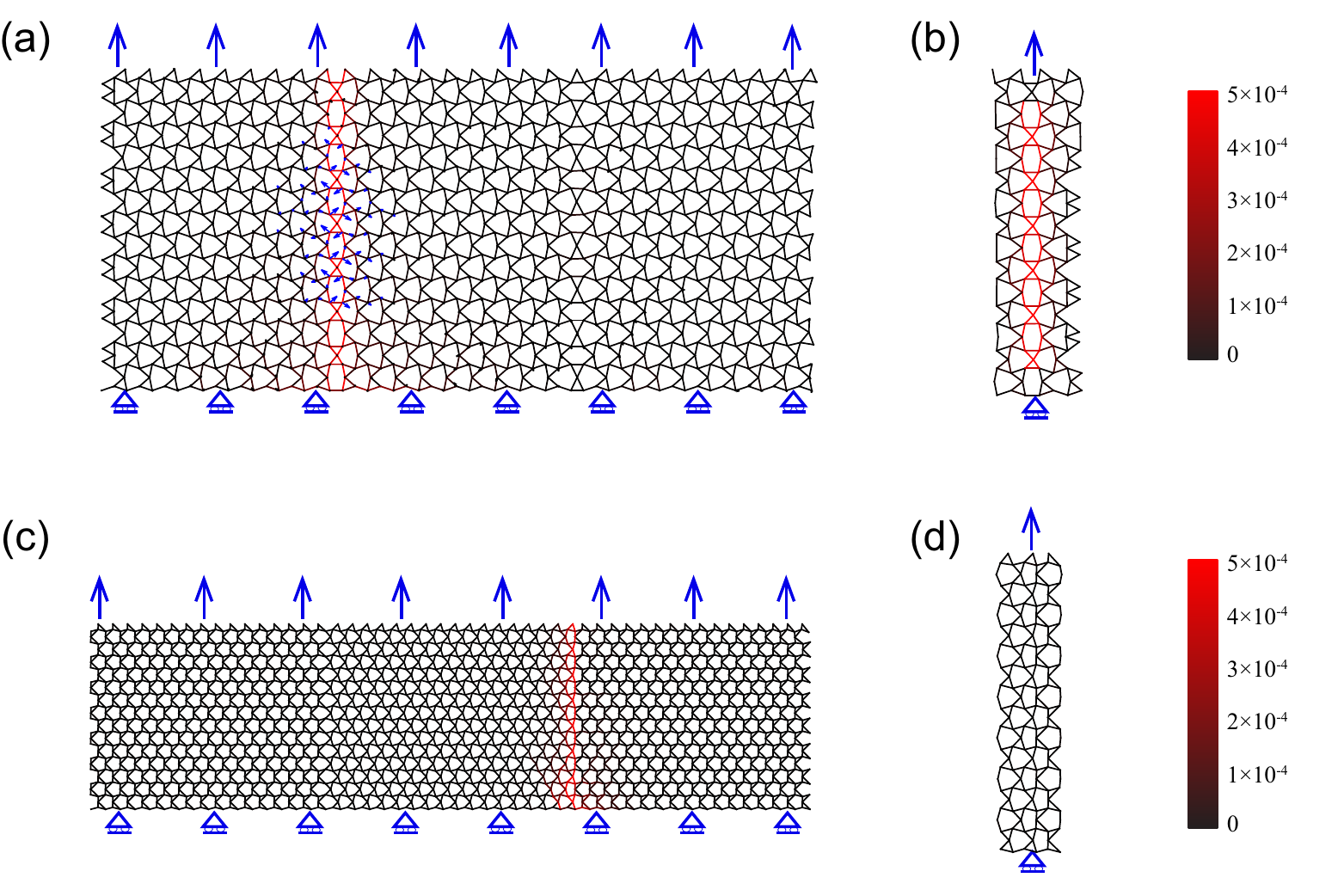}
\caption{Comparison between localized states of self stress that are non-topological and topological.  
(a) A non-topological lattice ($\bm R_{\rm T}=0$ in all domains) that exhibits localized states of self stress. The lattice contains three domains of twisted kagome lattices of alternating twisting directions with two domain walls.  Any sub region of this lattice must have $N_{SS}=N_0$.  
(b) A thin layer of lattice that is cut out of lattice (a) to isolate the states of self stress,  showing that the localized states of self stress in lattice (a) isn't a result of phonon band topology. Because $N_{SS}=N_0$ in this strip, there must be corresponding localized floppy modes.  One example of such floppy modes is shown by arrows in (a).  
(c) A topological deformed kagome lattice ($\bm R_{\rm T}\neq 0$) which has a domain wall of $\nu^{\mathcal{S}}<0$ and localized states of self stress that are protected by the bulk phonon band topology. 
(d) The thin strip that contains the state of self stress domain wall is cut off from the lattice in (c).  Without the bulk, the state of self stress is no longer present in the cut-off strip.
}
\label{fig4}
\end{center}
\end{figure}

It is worth noting that localized stress distributions that appear to be similar to those shown in fig.\ref{fig3} can also emerge without a topological origin. 
An example of this is shown in fig.~\ref{fig4}(a), where we show a lattice with $\bm{R}_{\rm T}=0$ in all domains but still shows localized states of self stress.  The structure contains twisted kagome lattices~\cite{sun2012surface} of opposite direction of twisting in its neighboring domains.  In particular, the pointing up triangles rotate clockwise in the two domains on the left and right and rotate counterclockwise in the middle domain, leading to two domains walls.  There is stress localization at the left domain wall.  
This, however, does not conflict with the theory of topological mechanics. In this structure, the left domain wall has topological charge $\nu^\mathcal{S}=0$, which indicates equal numbers of floppy modes $N_0$ and states of self stress $N_{SS}$. 
As long as $N_0=N_{SS}$, they can both localize at the domain wall.  Indeed, our calculation through the null spaces of matrix $C$ and $Q$ show pairs of states of self stress and floppy modes at this domain wall.  
We marked one of the floppy modes in fig.\ref{fig4}(a) with blue arrows in the kagome lattice. 
To explicitly show that the nonzero $N_{SS}$ and $N_0$ here are not a consequence of the lattice topology, but rather the local geometry of the lattice in the region, we cut a thin strip of the lattice that contains the localized states of self stress but doesn't include the bulk of the lattice.
As shown in fig.\ref{fig4}(b), the linear response of the thin strip still contains the localized states of self stress, despite the fact that it is cut from the lattice bulk.

For comparison, we take the deformed kagome lattice shown in fig.\ref{fig3}(c) that has nonzero $\bm R_{\rm T}$ and a domain wall of $\nu^{\mathcal{S}}<0$, and cut a thin strip containing this domain wall out of the lattice bulk. We then test the linear response of the thin strip alone, which is shown in fig.~\ref{fig3}(d). 
Interestingly, the states of self stress in the thin strip disappear when it's cut off from the lattice bulk.  This is consistent with the theory of topological mechanics that the topologically protected states of self stress comes from the bulk phonon band topology, and when the domain wall is isolated out the exponentially localized states of self stress no longer exist.  A similar demonstration for finite-frequency edge states has been shown in Ref.~\cite{nash2015topological} in a gyroscopic system.

\begin{figure}[h]
\begin{center}
\includegraphics{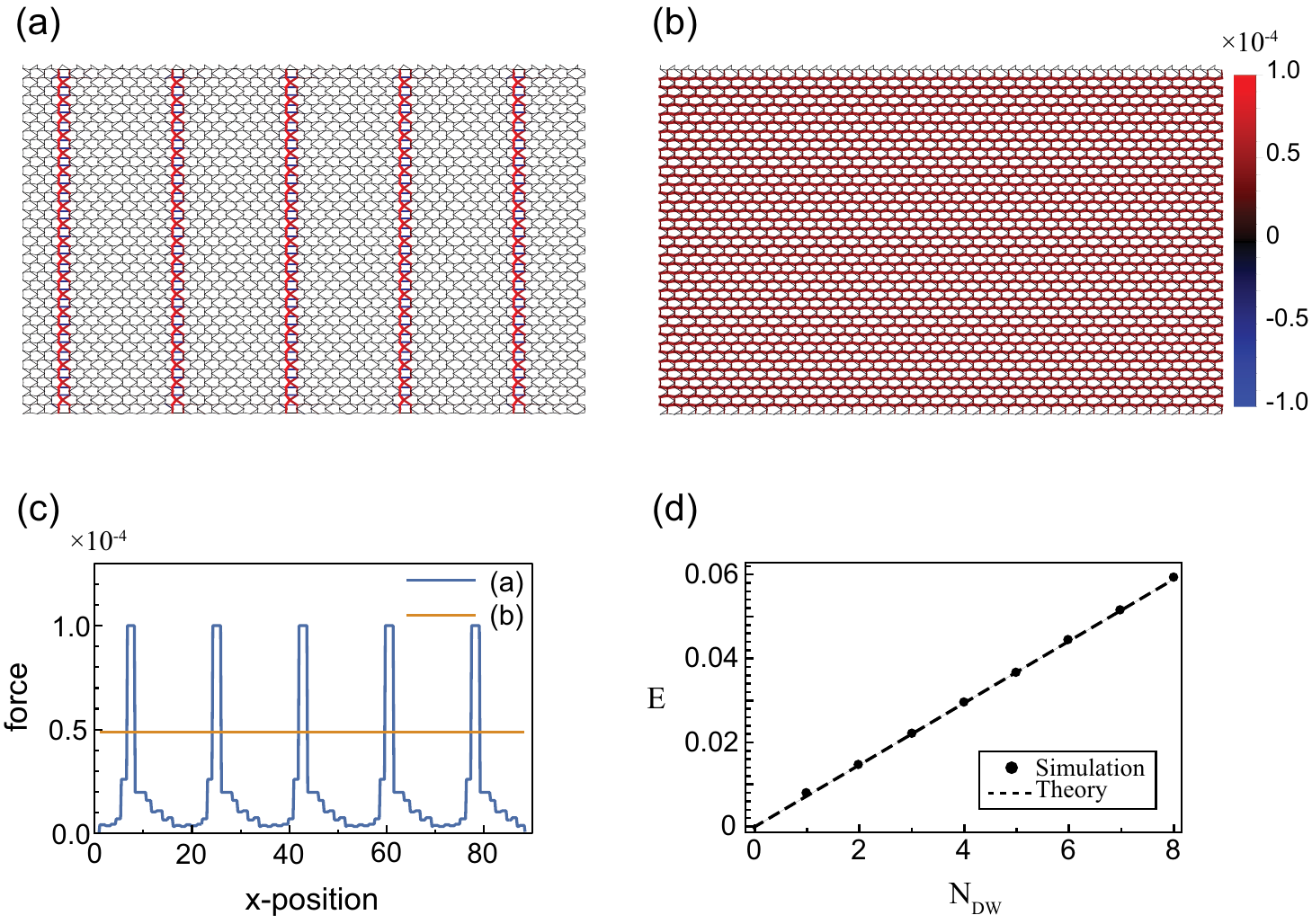}
\caption{
(a) A deformed kagome lattice with domains of $\bm R_{\rm T}$ pointing to left and right in alternating order, creating self-stress and floppy-mode domain walls.  (b) A deformed kagome lattice of homogeneous geometry and $\bm R_{\rm T}$ pointing to the left with no domain walls.  In (a) and (b) the lattices are stretched in the vertical direction with the same force, and the bonds are colored according to tension they carry [legend at the right of (b)].  (c) Bond forces in the lattices in (a) and (b).  At each horizontal ($x$) position, we check all bonds that cross this horizontal coordinate and plot the tension of the most stretched bond.  Under the same external load, bond forces are much lower in the bulk of the deformed kagome lattice with domain walls.  
(d) The elastic modulus $E$ against uniaxial stretching in the vertical direction of lattices as shown in (a) as a function of the number of domain walls $N_{DW}$. 
 }
\label{fig5}
\end{center}
\end{figure}

\begin{figure}[htbp]
\begin{center}
\includegraphics{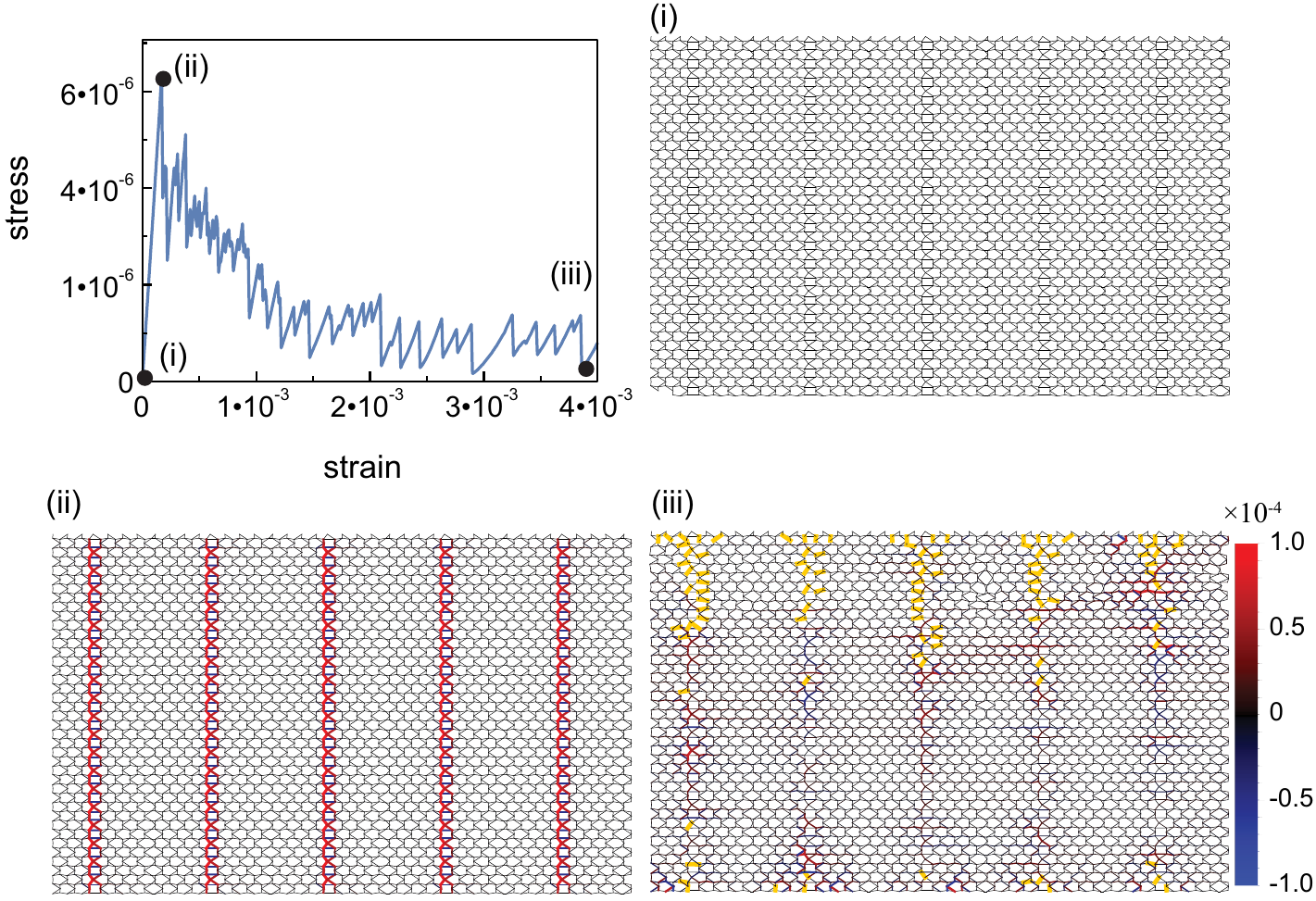}
\caption{
(a) The stress-strain curve for a deformed kagome lattice with multiple self-stress domain walls during the fracturing process. The initial point (i), peak point (ii), and one of the valley points (iii) are marked on the curve, with snapshots of the system at these points shown. Bonds are colored according to the tension they bear (legend of tension scale at the lower right panel).  
In (iii), broken springs that are removed from the lattice is colored yellow. 
}
\label{fig7}
\end{center}
\end{figure}

Next, we test the effect of embedding multiple domain walls in the lattices.  Topologically protected states of self stress are characterized by their decay rate $\kappa$ as we discussed in Sec.~\ref{SEC:Theory}.  Only when the bulk domain depth is greater than $1/\kappa$, the state of self stress is well defined to be exponentially localized.  
Therefore, two domain walls that are separated by a distance $\Delta w\gg 2/\kappa$ can be treated as independent. This allows us to design $N_{DW}$ domain walls that are independent to each other in a single lattice provided that the lattice depth $w>N_{DW} \Delta w$.

The advantages of designing mechanical metamaterials with multiple self-stress domain walls are the following.  First, the self-stress domain walls ``attract'' stress and protect regions in between.  This is shown in fig.~\ref{fig5}(a-c) where we compare a lattice with multiple self-stress domain walls and a deformed kagome lattice with the same type of unit-cell geometry but only one domain.  Under the same vertical tension, the lattice with domain walls only have high stress at the domain walls and the regions in between bear very low stress, whereas the one-domain lattice is homogeneously stressed.  In fig.~\ref{fig5}(c) we show quantitatively the bond force comparison between the two lattices.  
Second, by controlling the density of these domain walls we can also control the elastic modulus, $E\propto N_{DW}$, under the condition that domain walls are independent, as shown in eq.~(\ref{tension}) and in fig.~\ref{fig5}(d).  Third, because the exponentially localized states of self stress are topologically protected, as we discussed above, they continue to attract stress deep into the fracturing process, leading to a gradual failure, in contrast to catastrophic failure in conventional brittle materials (fig.\ref{fig7}).  Videos of our simulations of the fracturing process of Maxwell lattices with one or multiple domain walls, as well as brittle failure of the regular kagome lattice are included in the Supplementary Materials.

\section{Conclusion}
To summarize, we investigated how Maxwell lattices with domain walls hosting topologically protected states of self stress and floppy modes fracture under applied stress.  We find that bond breaking events concentrates near self-stress domain walls, providing protection to the lattice bulk, even deep into the failure process.  

Our results open the door to the design of high strength mechanical metamaterials based on topological mechanics.  By controlling the line density of the domain walls, we can control both the elastic moduli and the fracturing process of the structure.  We show that in the presence of self-stress domain walls, the bulk of the lattice is protected from fracturing, even when small cracks exist in the bulk.  This is a useful property that can be used in structures where perfect periodicity in the bulk needs to be protected for functions, such as wave-manipulating acoustic metamaterials.

\section{Acknowledgements}
The authors gratefully acknowledge supports from the National Science Foundation Grant No. NSF DMR - 1609051 and NSF EFMA - 1741618.

\section*{References}

%\bibliography{bib}
\bibliographystyle{unsrt}

\newpage
\appendix
\section*{Appendix}
\setcounter{section}{1}

In this Appendix, we show the derivation of eq.~(\ref{tension}) for the spring tensions $\bm t$ in the bulk of the lattice. It has been shown in the main text that $\bm t$ must be a superposition of states of self stress in the bulk, i.e. 
\begin{equation}
\bm t = \sum_i a_i \bm s_i,\label{linear}
\end{equation}
where $\{\bm s_i\}$ is the complete orthonormal basis of the null space of $Q_V$. Using $\bm t = K \bm  e$ we can write down the explicit expression for the coefficients $a_i$ as
\begin{eqnarray}
a_i = \bm s_i \cdot \bm t = \bm s_i \cdot K \bm e,\\
0 = \bm r_i \cdot \bm t = \bm r_i \cdot K \bm e,
\end{eqnarray}
where $\{\bm r_i\}$ are the set of eigenvectors for nonzero eigenvalues of matrix $Q_V$. 
We then plug in eq.~(\ref{extension}) in the main text to rewrite $\bm e$ using $\bm u_V$ and $\bm u_{\partial V}$, i.e. 
\begin{eqnarray}
a_i = \bm s_i \cdot K (C_V \bm u_V + C_{\partial V} \bm u_{\partial V}),\\
0 = \bm r_i \cdot K (C_V \bm u_V + C_{\partial V} \bm u_{\partial V}).
\end{eqnarray}

This can be simplified by decomposing the spring constant matrix $K$ into the null space and the orthogonal complement space of $Q_V$ into
\begin{equation}
K = 
\left(\begin{array}{cc} K_{rr} & K_{rs}\\ K_{sr} & K_{ss} \end{array}\right),
\end{equation}
where $(K_{rr})_{ij} \equiv \bm r_i \cdot K \bm r_j$, $(K_{rs})_{ij} = (K_{sr})_{ji}\equiv \bm r_i \cdot K \bm s_j$ and $(K_{ss})_{ij} \equiv \bm s_i \cdot K \bm s_j$. Note that $K_{rr}$ is invertible because its eigenvalues are nonzero. 
We then simplify our expression by denoting $a_i$ with a vector $\bm a$ such that $(\bm a)_i \equiv a_i$, and have the equations for the linear coefficients as
\begin{eqnarray}
\bm a = K_{sr} P_r C_V \bm u_V + (K_{sr} P_r+K_{ss}P_s)C_{\partial V} \bm u_{\partial V},\label{coeff}\\
0 = K_{rr} P_r C_V \bm u_V +(K_{rr}P_r +K_{rs}P_s)C_{\partial V} \bm u_{\partial V},\label{balance}
\end{eqnarray}
where $P_s$ is the projection operator from the original bond label space into the states-of-self-stress space (null space of $Q_V$), and $P_r$ is the projection operator from the original bond label space into the orthogonal complement space.  We have also used the property that $C_V \bm s_i=0$ for all states of self stress $i$.  

From eq.~(\ref{balance}), we  obtain 
\begin{equation}
P_r C_V \bm u_V = - K_{rr}^{-1}(K_{rr} P_r +K_{rs} P_s) C_{\partial V} \bm u_{\partial V}.
\end{equation}
Finally, by plugging this relation into eq.~(\ref{coeff}), we obtain our linear coefficients
\begin{equation}
\bm a = \left[-K_{sr}K_{rr}^{-1}(K_{rr} P_r + K_{rs}P_s) + K_{sr}P_r +K_{ss}P_s\right]C_{\partial V} \bm u_{\partial V},
\end{equation}
which can be further simplified as
\begin{equation}
\bm a = (K^{-1})_{ss} P_s C_{\partial V} \bm u_{\partial V},
\end{equation}
using a matrix identity that $(K^{-1})_{ss} = K_{ss} - K_{sr}(K_{rr}^{-1})K_{rs}$. Putting the coefficient back into eq.~(\ref{linear}), we arrive at eq.~(\ref{tension}) in the main text.

\end{document}